# Physics-Inspired Regularized Pulse-Echo Quantitative Ultrasound: Efficient Optimization with ADMM

Noushin Jafarpisheh[1], Laura Castaneda Martinez[2], Hayley Whitson[3], Ivan M. Rosado-Mendez[4], and Hassan Rivaz[5]

*Abstract*— **Pulse-echo Quantitative ultrasound (PEQUS), which estimates the quantitative properties of tissue microstructure, entails estimating the average attenuation and the backscatter coefficient (BSC). Growing recent research has focused on the regularized estimation of these parameters. Herein, we make two contributions to this field: First, we consider the physics of the average attenuation and backscattering to devise regularization terms accordingly. More specifically, since the average attenuation gradually alters in different parts of the tissue while BSC can vary markedly from tissue to tissue, we apply *L2* and *L1* norms for the average attenuation and the BSC, respectively. Second, we multiply different frequencies and depths of the power spectra with different weights according to their noise levels. Our rationale is that the high-frequency contents of the power spectra at deep regions have a low signal-to-noise ratio. We exploit the alternating direction method of multipliers (ADMM) for optimizing the cost function. Qualitative and quantitative evaluation of bias and variance exhibit that our proposed algorithm substantially improves the estimations of the average attenuation and the BSC.**

## I. INTRODUCTION

Ultrasound imaging (US) has numerous medical applications [1]–[3]. Albeit real-time, cost-effective, and portable, it depicts a qualitative map of underlying tissue that can be affected by the settings of an ultrasound machine. As a result, its interpretation is contingent on the operator's skills. Pulse-echo quantitative ultrasound (PEQUS) resolves the drawbacks mentioned above by revealing the quantitative properties of tissue microstructure [4]–[7]. Fatty liver classification, [8], [9], prostate cancer imaging [10], bone assessment [11], and breast cancer monitoring [12] are a few examples of clinical applications of PEQUS. PEQUS techniques are categorized into different fields [13]–[17]. Spectral-based techniques [18]–[20] delve into the power spectra of the radio frequency (RF) data backscattered from tissue. These techniques provide estimations of acoustic properties of tissue which comprise attenuation [21]–[24], backscatter coefficient (BSC) [25], and scatterer properties [26]–[28]. Herein, we focus on attenuation and BSC estimation.

Previous studies [29]–[34] reported regularized-based techniques exceedingly outperform the non-regularized schemes. Vajihi *et al.* [31], [32] introduced dynamic programming (DP) for BSC estimation with implicit compensation for the total attenuation, i.e., the attenuation produced by tissues between the transducer and the region of estimation. However, a Matlab implementation of DP indicates it requires above 2000 seconds and therefore, it is not practical for clinical purposes. Destrempes *et al.* [35], optimized a generalized least absolute shrinkage and selection operator (LASSO) problem using a Lagrangian multiplier in addition to the Bayesian Information Criterion for estimating attenuation and scatterer properties. Deeba *et al.* [36] presented a cost function entailing total variation regularization to estimate effective scatterer diameter (ESD) and acoustic concentration (AC) solved using the CVX toolbox [37] in MATLAB. Rafati *et al.* [38] aimed at estimating attenuation coefficient map using the attenuation coefficient slope. Accordingly, they approximated the logarithmic scale of the power spectra at each frequency as a function of depth. Consequently, the frequency where the maximum y-intercept occurred was selected for calculating the local attenuation coefficient slope. Afterward, the frequency range was limited to those where the ratio of the normalized power spectra with respect to the depths at two locations and two adjacent frequencies was within 25% of the estimated local attenuation coefficient slope. The ultimate estimate of the local attenuation coefficient slope was calculated using a linear regression on the confined frequency range. Birdi *et al.* [39] introduced an analytical regularized based method to directly estimate local attenuation providing a physical-based model. Additionally, performing this technique prevents post-processing computation for mapping effective attenuation to the local attenuation. Furthermore, to exploit the advantages of the *L1* norm, the authors proposed a weighted *L2* norm scheme that can iteratively approximate the *L1* norm. Our group recently proposed a novel analytical technique called AnaLytical Globally rEgularized BacksccatteR quAntitative ultrasound, or ALGEBRA [29], to estimate the total attenuation and the BSC. ALGEBRA optimizes a penalty function containing the *L2* norm in both data and regularization terms. However, it suffers from two major issues as follows:

1) ALGEBRA considers *L2* norm regularization for both the average attenuation from intervening tissues (hereforth referred to as $\alpha_{avg}$), and the BSC. Nonetheless, according to the physics, the $\alpha_{avg}$, and BSC have different rates of spatial variations (due to the averaging effect in $\alpha_{avg}$ and the different physical mechanisms underlying absorption and scattering). In general, BSC can change from tissue to tissue, while $\alpha_{avg}$, presents more gradual changes.

2) ALGEBRA considers equal involvement for power spectra at each frequency and depth in parameter estimation. However, different frequency components of the power spectrum present varying levels of signal-to-noise ratio (SNR) as a function of depth. A maximum likelihood estimator should put less trust in low SNR data.

[1]N. Jafarpisheh is with the Department of Electrical and Computer Engineering and PERFORM Center, Concordia University, Canada.
 n_jafarp@encs.concordia.ca.
[2] Laura C. Martinez is with the Department of Medical Physics, University of Wisconsin-Madison, Madison, WI, USA.
 castanedamar@wisc.edu.
[3]Hayley Whitson is with the Department of Medical Physics, University of Wisconsin-Madison, Madison, WI, USA
 hwhitson@wisc.edu.
[4]Ivan M. Rosado-Mendez is with the Departments of Medical Physics and Radiology, University of Wisconsin-Madison, Madison, WI, USA.
 rosadomendez@wisc.edu.
[5]H. Rivaz is with the Department of Electrical and Computer Engineering and PERFORM Center, Concordia University, Canada.
 hrivaz@ece.concordia.ca.

To overcome the drawbacks mentioned above, we utilize the *L2* norm for the average attenuation and the *L1* norm for the BSC in the regularization part of the cost function. To optimize the proposed cost function efficiently, we employ the Alternation Direction Method of Multipliers, or ADMM. ADMM has been previously used by Coila et al. [40] for the regularization of local attenuation estimation. However, the authors did not parameterize BSC through the power-law model. We also introduce a criterion to define weights that vary with frequency and depth.

In this report, we make the following contributions to the field of regularized PEQUS: 1) we incorporate physics-based regularization norms for attenuation and BSC; 2) we applying ADMM to penalize the cost function instead of the CVX toolbox in MATLAB used by Deeba et al. [36], [41]; 3) we assign SNR-based weights that vary with depth and frequency to the echo signal power spectrum;. 4) In contrast to [36], we estimate $\alpha_{avg}$, and BSC instead of ESD and AC.

Our proposed method is explained in Section II. Section III presents the results, followed by a discussion part in Section IV. Conclusions are laid out in Section V.

## II. METHODS

Similar to [29], [31], [42], we start with the reference phantom method (RPM) [43], a well-known approach employed to cancel system dependencies.

In RPM, the power spectra of a sample phantom *s* with unknown PEQUS parameters $\alpha_{avg}$ and BSC, represented by $\sigma_b$, are normalized by the power spectra of a calibrated reference phantom *r*. Both phantoms are scanned with the same transducer, sound speed, and machine setting:

$$\frac{S_s(f;z,x)}{S_r(f;z)} = \frac{\sigma_{b,s}(f;z,x)A_s(f;z,x)}{\sigma_{b,r}(f)A_r(f,z)} \quad (1)$$

where $\sigma_b(f;z)$, $A(f;z)$ entitled BSC and total attenuation are defined through a power-law parametrization and an exponential function, respectively:

$$\sigma_b(f;z) = \beta(z)f^{\nu(z)} \quad (2)$$
$$A(f;z) = exp(-4f\alpha_{avg}z) \quad (3)$$

where *f*, and *z* refer to the frequency, and depth. More information can be found in [29]. Taking the natural log from both sides of Eq. (1) and applying $\sigma_b(f;z)$ and $A(f;z)$ in Eqs. (2, 3) for sample and reference phantom, the linear equation is obtained:

$$X(f;z) = -4a(z)fz + b(z) + n(z)\ln f \quad (4)$$

where:
$$a(z) = \alpha_{eff}(z) - \alpha_{0,r}$$
$$b(z) = \ln\beta_s(z) - \ln\beta_r,$$
$$n(z) = \nu_s(z) - \nu_r.$$

Our proposed method is based on minimizing a cost function *C* composed of a data term *D* and a regularization term *R*. Herein, the data term is formulated as presented in [29]. The regularization term will be presented in the next part.

$$C = D + R \quad (5).$$

$$D = \sum_{l=1}^{N_F}\sum_{i=1}^{N_R}(X(f_l,z_i) - b_i - n_i\ln(f_l) + 4a_if_lz_i)^2 \quad (6)$$

where $N_R$ and $N_F$ refer to the number of axial lines and frequency bins, respectively.

### A. ADMM for L1 norm regularization in QUS

We consider two different regularization functions $R_1$ and $R_2$ as follows.

In $R_1$, we use the *L1* norm regularization for all the parameters:

$$R_1 = \sum_{i=2}^{N_R} w_a|a_i - a_{i-1}| + w_b|b_i - b_{i-1}| + w_n|n_i - n_{i-1}| \quad (7)$$

where $w_a$, $w_b$, and $w_n$ refer to the regularization weights for each parameter.

In $R_2$, we use *L2* norm for the average attenuation and *L1* for the BSC-related terms:

$$R_2 = \sum_{i=2}^{N_R} w_a(a_i - a_{i-1})^2 + w_b|b_i - b_{i-1}|$$
$$+ w_n|n_i - n_{i-1}| \quad (8).$$

Here, the issue is that *L1* norm is not analytically differentiable. To solve this, consider the goal is minimizing the following constrain cost function:

$$C = D(x) + R(s) \quad (9)$$
$$\text{subject to } Kx + Ls = m$$
where $K$ and $L$ are known matrixes and $m$ is a given vector, $x$ and $s$ are separable variables.

The augmented Lagrangian function solves the unconstrained version of the above constrained cost function as follows:

$$L\rho(x, s, y) = D(x) + R(s) + y^T(Kx + Ls - m) + \left(\frac{\rho}{2}\right)||Kx + Ls - m||_2^2 \quad (10)$$

where $y$ is the Lagrange multiplier and $\rho > 0$ weights the constrain.

ADMM solves the Eq. (9) iteratively as follows:
$$x^{k+1} = \text{argmin}_x(L\rho(x, s^k, y^k))$$
$$s^{k+1} = \text{argmin}_s(L\rho(x^{k+1}, s, y^k))$$
$$y^{k+1} = y^k + \rho(Kx^{k+1} + Ls^{k+1} - m) \quad (11)$$

where $k$ shows the iteration number.

Therefore, as ADMM's name implies, it penalizes each component of the cost function sequentially (alternative direction). This feature is favorable when an identical optimization procedure is not applicable for both data and regularization terms.

Herein, we start with rewriting Eq. (6) in the matrix format and call the obtained equation $Qx = Y$ where $x = [a_1 \, a_2 \dots a_{N_R} \, b_1 \dots b_{N_R} \, n_1 \dots n_{N_R}]^T$, $Q$ is a matrix containing depth and frequency and $Y$ is a vector. We multiply both sides of this equation by $Q^T$ and name $Q^TQ = H$, $Q^TY = t$, and minimize the following cost function:

$$C = \frac{1}{2}||Hx - t||_2^2 + \lambda ||s||_1 \quad (12)$$
$$\text{subject to } Kx - s = 0$$

where $s$ denotes the regularization term and $H$ is given as:

$$H = \begin{bmatrix} H_1 & H_2 & H_3 \\ H_2 & H_4 & H_5 \\ H_3 & H_5 & H_6 \end{bmatrix} \quad (13)$$

and $H_j, j = 1, \dots, 6$, are $N_R \times N_R$ diagonal matrixes:

$$H_1 = \left(16 \sum_{l=1}^{N_F} f_l^2\right) Z_2, \quad H_2 = \left(-4 \sum_{l=1}^{N_F} f_l\right) Z_1,$$
$$H_3 = \left(-4 \sum_{l=1}^{N_F} f_l \ln f_l\right) Z_1, \quad H_4 = (N_F)I,$$
$$H_5 = \left(\sum_{l=1}^{N_F} \ln f_l\right) I, \quad H_6 = \left(\sum_{l=1}^{N_F} (\ln f_l)^2\right) \quad (14)$$

where $I$ is the $N_R \times N_R$ identity matrix and

$$Z_1 = \begin{bmatrix} z_1 & 0 & \dots & 0 \\ 0 & z_2 & \dots & 0 \\ \vdots & \vdots & \ddots & \vdots \\ 0 & 0 & \dots & z_{N_R} \end{bmatrix}, \quad Z_2 = \begin{bmatrix} z_1^2 & 0 & \dots & 0 \\ 0 & z_2^2 & \dots & 0 \\ \vdots & \vdots & \ddots & \vdots \\ 0 & 0 & \dots & z_{N_R}^2 \end{bmatrix}$$

and $t$ is defined as:

$$t = \begin{bmatrix} t1 \\ t2 \\ t3 \end{bmatrix}, \quad (15)$$

where its $i^{th}$ term is:
$$t1_i = -4z_i \sum_{l=1}^{N_F} X(f_l, z_i) f_l,$$
$$t2_i = \sum_{l=1}^{N_F} X(f_l, z_i),$$
$$t3_i = \sum_{l=1}^{N_F} X(f_l, z_i) \ln f_l$$

and $K$ is:

$$K = \begin{bmatrix} K_a & 0 & 0 \\ 0 & K_b & 0 \\ 0 & 0 & K_n \end{bmatrix}$$

and
$$K_j = w_j B \quad (16)$$

where $j$ can be $a$, $b$, $n$, and $w_j$ denotes the regularization weights for the corresponding parameter and $B$ is:

$$B = \begin{bmatrix} 1 & -1 & 0 & 0 & \cdots & 0 \\ 0 & 1 & -1 & 0 & \cdots & 0 \\ \vdots & \vdots & \vdots & \vdots & \ddots & \vdots \\ 0 & 0 & 0 & \cdots & 1 & -1 \end{bmatrix} \quad (17)$$

Now, to update $x$, $s$, and $y$, we exploit Eqs. (10), (11), where $D(x)$ and $R(s)$ are $\frac{1}{2}||Hx - t||_2^2$ and $\lambda ||s||_1$, respectively. Accordingly, we obtain:

$$x^{k+1} := (H^TH + \rho K^TK)^{-1} H^T t + \rho K^T (s^k - y^k))$$

To update $s$, the following equation is obtained with respect to the shrinkage function $S_{\frac{\lambda}{\rho}}$:

$$s^{k+1} = \text{argmin}_s \lambda ||s||_1 - y + \frac{\rho}{2}||Kx^{k+1} - s||_2$$

which performs as soft thresholding:
$$S_{\frac{\lambda}{\rho}} = \text{sgn}(\cdot) \max\{|\cdot| - \frac{\lambda}{\rho}, 0\}$$

Consequently, we have:
$$s^{k+1} := S_{\frac{\lambda}{\rho}}(Kx^{k+1} + y^k)$$
$$y^{k+1} = y^k + Kx^{k+1} - s^{k+1} \quad (18).$$

In contrast to the BSC, which can vary abruptly between different tissue regions, the average attenuation varies moderately. To be consistent with this physical property, we propose employing the *L2* norm for the average attenuation and the *L1* norm for the BSC in the regularization part instead of incorporating the same norms for all the parameters. Accordingly, we split the vector $x$ into two parts for the average attenuation $x_1$ and the BSC $x_2$. Consequently, the corresponding elements of $s$ and $y$ are called $s_1, s_2, y_1, y_2$. Therefore, the algorithm is modified as follows:

$$C = \frac{1}{2}||Hx - t||_2^2 + \lambda_1 ||s_1||_2^2 + \lambda_2 ||s_2||_1 \quad (19)$$
$$\text{subject to } K_1 x_1 - s_1 = 0, K_2 x_2 - s_2 = 0$$

Updating $x$ and $y$ are same as Eq. (18), but $s$ is updated by $s_1$ and $s_2$ as follows:

$$x^{k+1} := (H^T H + \rho K^T K)^{-1} H^T t + \rho K^T (s^k - y^k)$$

$$s_1^{k+1} := (K_1 x_1^{k+1} + y_1^k)/(\rho + \lambda 1)$$

$$s_2^{k+1} := S_{\frac{\lambda_2}{\rho}}(K_2 x_2^{k+1} + y_2^k)$$

$$y^{k+1} = y^k + K x^{k+1} - s^{k+1} \quad (20)$$

where $H$ and $K$ are same as Eqs. (13) and (16), and $K_1, K_2, x_1, x_2$ are:

$$K_1 = K_a, K_2 = \begin{bmatrix} K_b & 0 \\ 0 & K_n \end{bmatrix} \quad (21)$$

$$x_1 = [a_1\ a_2\ ...\ a_{N_R}]^T, x_2 = [b_1\ ...\ b_{N_R}\ n_1\ ...\ n_{N_R}]^T \quad (22)$$

*B. Weighted Frequency*

We use a weighted frequency scheme to give the power spectra a score (weight) at each frequency and depth. To do so, the data term is multiplied by $w_d$. Thus, the data term formulation is adapted as follows:

$$D = \sum_{l=1}^{N_F} \sum_{i=1}^{N_R} w_d(l,i)(X(f_l, z_i) - b_i - n_i \ln(f_l) + 4a_i f_l z_i)^2 \quad (23).$$

To select the bandwidth, we plot the logarithmic scale of the lateral average of the power spectra of sample and reference phantoms and determine the corresponding frequency range where higher power is amassed. The black dash lines in Figs. 1 and 2 illustrate this range for example power spectra obtained from a sample phantom and a reference phantom, which are the corresponding frequencies for 80% of the maximum logarithmic power spectra. Herein, the background of the sample phantom is the reference phantom. To calculate $w_d$, we propose using contour level sets depicted in Figs. 3 and 4, and to carry out this, we form the following equations to compute $w_s$ and $w_r$ for the sample and reference phantoms separately.

$$w_s = \begin{cases} 1, & S_s > T_{1s} \\ \frac{S_s - T_{2s}}{T_{1s} - T_{2s}}, & T_{1s} > S_s > T_{2s} \end{cases}$$

$$w_r = \begin{cases} 1, & S_r > T_{1r} \\ \frac{S_r - T_{2r}}{T_{1r} - T_{2r}}, & T_{2s} > S_r > T_{2r} \end{cases}$$

where $S$ is the logarithmic scale of the power spectra shown in Figs. 1 and 2, and $T_1$ and $T_2$ express the upper level and lower level of contours shown in Figs. 3 and 4 with yellow and purple colors, respectively. The upper and lower level are 90% of the maximum and 167% of the minimum of the logarithmic power spectra.

In the end, the intersection of the $w_s$ and $w_r$ computed using element-wise matrix multiplication is regarded as $w_d$:

$$w_d = w_s \odot w_r \quad (24).$$

Fig. 1(a) exhibits power spectra of the sample phantom that were affected by the specular reflectors. Accordingly, we only consider the reference phantom to calculate $w_d$ ($w_d = w_r$).

Figs. 5 and 6 demonstrate $w_d$ for the region with specular reflectors and the region with inclusions.

*C. Tissue mimicking phantom and data acquisition*

Two regions of a phantom named Gammex 410SCG phantom (Gammex-Sun Nuclear, Middleton, WI) was scanned with following properties. In both regions, the background is the reference phantom.

*1) Region with specular reflectors*

Five uncorrelated frames of RF data were collected, using the same transducer and operation parameters as the Gammex 410 SCG phantom. This region contains the presence of three nylon filaments of 0.1 mm in diameter. To get the Nylon filaments to produce specular reflectors, they were scanned placing the transducer perpendicular to the front face of the phantom. These structures produce echo signals originated from a coherent scattering process, despite the fact that these structures can commonly be present prior to the region of interest or within it, they violate the assumption of incoherent scattering process on which the RPM is based, reducing the accuracy and precision of acoustic parameters such as the backscatter coefficient [44], [45]. Therefore, the importance of this analysis was to evaluate the susceptibility of the method in the estimation of QUS parameters, in terms of bias and variance, due to the presence of specular reflectors.

The background has the following properties:
- $\alpha_{eff} = 0.6035\ dB\ cm^{-1}\ MHz^{-1}$
- $\beta_r = 2.9966 \times 10^{-6}\ cm^{-1} sr^{-1} MHz^{-\nu}$
- $\nu_n = 3.4281$
- $\sigma_{b,r}(8\ \text{MHz}) = 3.74 \times 10^{-3}\ cm^{-1} sr^{-1}$

*2) Region with inclusions*

Ten uncorrelated frames of RF data were collected, using the homogeneous region as the reference with an L11-5v transducer operated at an 8 MHz center frequency on a Verasonics Vantage 128 system (Verasonics, Kirkland, WA).

This region is composed of three cylindrical inclusions with +12dB, +6dB, and -6dB scattering respecting the background.

The $\alpha_{avg}$ and the BSC of this region are same as the region with specular reflectors.

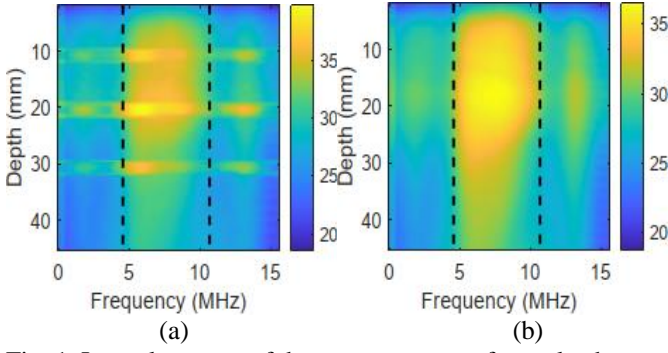

Fig. 1. Lateral average of the power spectra of sample phantom (a) and reference phantom (b) in the logarithmic scale for the region with specular reflectors.

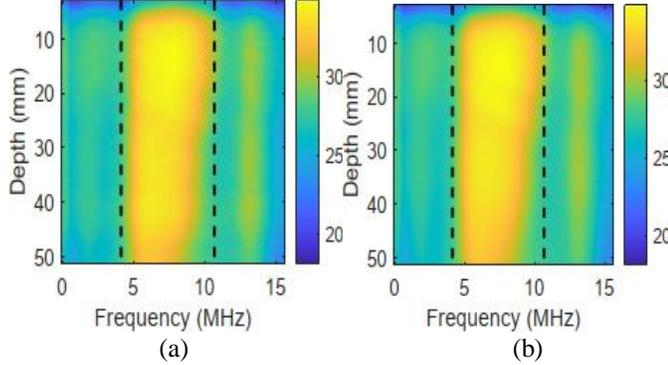

Fig. 2. Lateral average of the power spectra of sample phantom (a) and reference phantom (b) in the logarithmic scale for the region with inclusions.

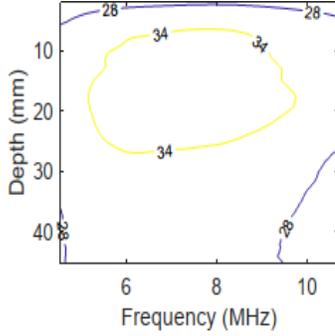

Fig. 3. Contour levels of the lateral average of the power spectra of reference phantom in the logarithmic scale for the region with specular reflectors.

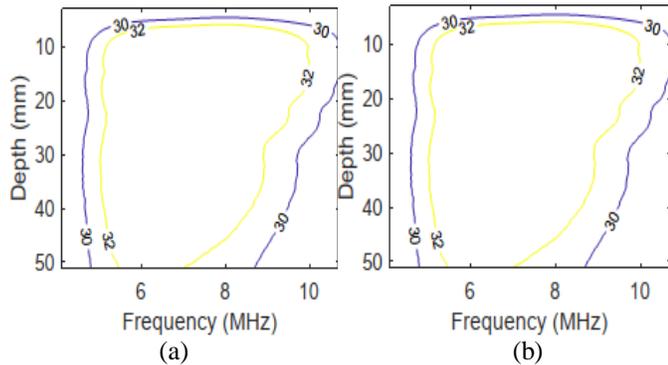

Fig. 4. Contour levels of the lateral average of the power spectra of sample phantom (a) and reference phantom (b) in the logarithmic scale for the region with inclusions.

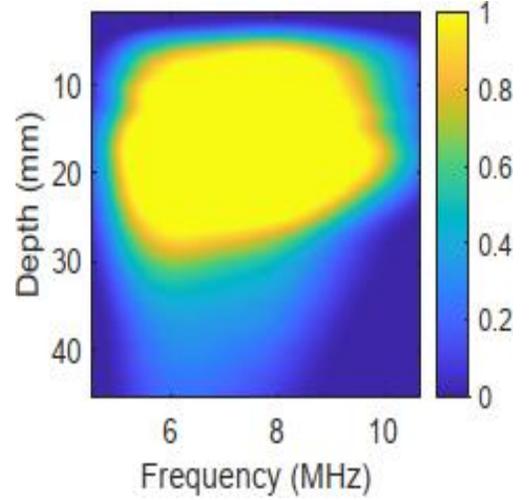

Fig. 5. $w_d$ for the region with specular reflectors.

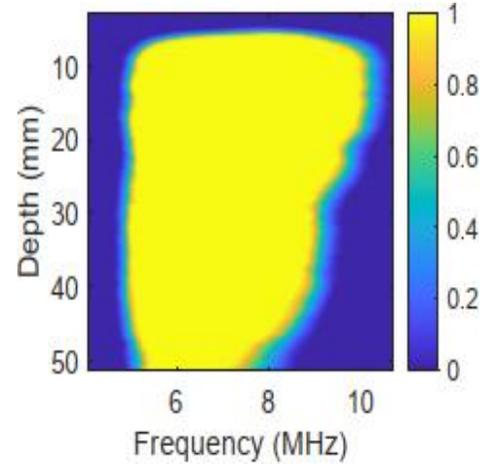

Fig. 6. $w_d$ for the region with inclusions.

*D. Quantitative metrics*

In this study, we report the variance and bias of the estimations to evaluate and compare the results. To compute the variance and bias, four regions of the interests (ROI) are selected for each region. Figs. 7 and 8 present the locations of each ROI on a B-mode image of the region with specular reflectors and the region with inclusion, respectively. For the region with specular reflectors the locations are selected between the fibers and for the region with inclusions at the center of the background and each inclusion.

To report the bias and variance for each ROI, we calculate the average of the estimations across the frames, $M$, and exploit the following equations. The BSC is scaled to dB w. r. to $10^{-4}$ as follows:

$$\text{bias\_}\alpha_{avg} = |M_{a_i}(:) - GT_a|$$
$$\text{variance\_}\alpha_{avg} = var(M_{a_i}(:))$$
$$\text{bias\_BSC} = 10\log_{10}\left(\frac{M_{BSC_i}(:)}{10^{-4}}\right) - 10\log_{10}\left(\frac{GT_{BSC}}{1e^{-4}}\right)$$
$$\text{variance\_BSC} = var(10\log_{10}\left(\frac{M_{BSC_i}(:)}{10^{-4}}\right))$$

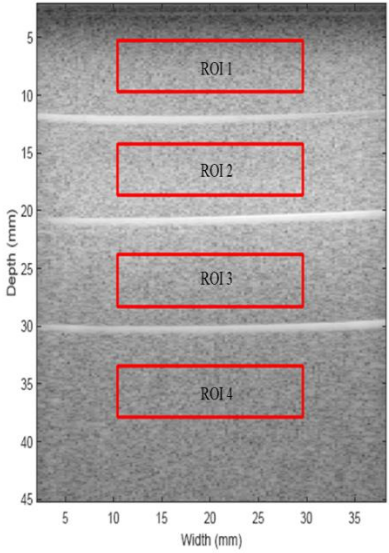

Fig 7. Location of ROI in the region with specular reflectors.

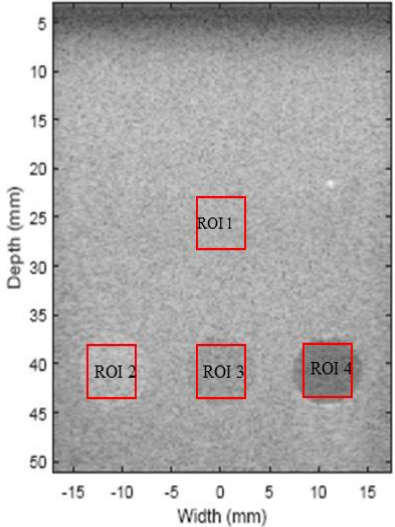

Fig 8. Location of ROI in the region with inclusions.

## III. RESULTS

Figs. 9 to 13 compare six techniques tested on the region with specular reflectors and the region with inclusions. Regularization weights are tuned individually for each set to ensure the best results are fulfilled. In real experiments where ground truths (GT) are unknown, choosing the weights can be more challenging. In such cases, one way to investigate what weights are proper is to first run least squares (LSQ) which is equal to setting zero weights for all the parameters and plot the QUS parameters. Then, we can set the weights to the values in the range of 0.1, 10, 1e2, …,1e8 and plot the results. Let us assume that for the weight 1e3, the results are almost identical to the LSQ. This means this weight is too small. Furthermore, if we suppose for the weight 1e7 the results are constant, it means the chosen weight is too large. In this case, 1e5 can be the optimal weight. To test this, we can double the optimal weight (2e5) and plot the results. If the results are similar to 1e5, then the optimal weight is 1e5. This implies that the results are independent of the weight.

### A. Region with Specular Reflector

Figs. 9 and 10 report the bias and variance of the BSC on the dB scale and the average attenuation. The Figs demonstrate the superior performance of ADMM in terms of bias and variance compared to the ALGEBRA. More precisely, comparing ADMM and ALGEBRA both without $w_d$ in regions 1 to 4 reveals the bias of BSC estimation decreased 86.1%, 88.3%, 76.3%, and 83.1%, respectively and the variance experienced 68.8%, 78.9%, 60.1% and 77.7% reduction. Moreover, the bias of average attenuation decreased 85.5%, 92.7%, 83%, and 93.2% in ROI 1 to 4 as well as 76.5%, 84.4%, 54.1%, and 76.2% reduction in the variance in each region is obtained. Additionally, associating ADMM with $w_d$, and different norms in the regularization term enhance the estimation.

Quantitative comparison of ALGEBRA with and without $w_d$ in Fig. 9 shows that smoother estimations are acquired by incorporating $w_d$ in the cost function. Moreover, engaging ADMM with $w_d$ is accompanied by the following quantitative analysis concerning not including $w_d$ in the cost function. Using the $w_d$ strategy in ADMM reduced the bias of BSC by 88.1%, 52.5%, and 98.9%, in regions 1 to 3, compared to not using $w_d$. Additionally, the variance is reduced by 42.1% in region 1. Moreover, attenuation evaluation results in 86.3%, 44.6%, decline in bias in regions 1 and 3. Regarding variance appraisal 52.2% lower variance is reported in region 1 when using ADMM with $w_d$ vs not using $w_d$. A Matlab implementation of ALGEBRA shows it is 100 times faster than ADMM for this region.

### B. Region with inclusion

The results of BSC estimation at center frequency using six techniques are depicted in Fig. 11. The visual comparison of the parametric images presents that ADMM surpasses ALGEBRA as it is more similar to the GT. From Figs. 12 and 13, it is perceived that encompassing $w_d$ in the ALGEBRA cost function reduces the bias and variance of the BSC and the average attenuation estimations. Comparing results of our proposed technique concerning ADMM without weighting the data term shows 80.3% reduction in the bias in region 2 as well as 15.5%, 22.1%, and 1% reduction in the variance of BSC estimation in region 1, 3, and 4, respectively. Furthermore, attenuation evaluation results in 29.4%, 88.2%, and 66.1% decline in the bias of regions 2 to 4 besides 4.2%, 10.9% decrement in the variance of regions 1 and 3. A Matlab implementation of ALGEBRA shows it is 18 times faster than ADMM for this region.
.

## IV. DISCUSSION

In this work, we exploited the RPM to formulate our proposed technique. Although new reference phantom free methods have been developed, the RPM can be feasibly implemented in the clinic without having to scan a phantom after each patient. Commercial scanners can be equipped by a pre-tuned reference phantom or data to train a machine learning model for providing the attenuation values. However, these systems provide the BSC measurement at one frequency [14].

This paper presented a novel penalty function consisting of *L2* norm weighted data term and *L1* and *L2* norm regularization terms optimized using ADMM. Our method was applied to data acquired from two regions of a phantom which we called the region with specular reflectors and the region with inclusions. Vajihi *et al.* [32] introduced applying *L1* norm in the cost function and penalized it using DP. However, DP suffers from disadvantages explained in detail in [29]. Two critical shortcomings related to DP's search-based strategy are the following:

1) DP is computationally burdensome, leading to not being capable for real-time functions.
2) In DP, a range for the GT should be known a priori, which may not be the case in real applications. If the search ranges do not contain the GT, the results will be incorrect.

Unlike [29], we carefully focused on selecting the bandwidth in the first part of our proposed technique. In [29], the frequency range was chosen based on the evaluation of the bias and variance of the BSC and the average attenuation estimations. Therefore, we observed for the region with inclusions a wide frequency range leads to the results that were in very good agreements with the GT. However, herein, our strategy for selecting the bandwidth of interest is different. Figs. 1 and 2 show that the frequency ranges were chosen based on the parametric image of the power spectra of the sample and reference phantoms. The intersection of the corresponding frequency ranges of the sample and the reference phantoms with high concentrated power is accounted for as the bandwidth of interest which is depth dependent as well. Our proposed approach for selecting the bandwidth of interest is more reasonable compared to the method we applied in [29] because of two points: 1) It is possible that a wide frequency range leads to excellent results in terms of the quantitative and qualitative assessments, but the selected band may contain low power, which is the case when sample and reference phantom are the same 2) In clinical applications with unknown GT, evaluation of bias which leads to selecting the frequency range is not practical.

In the next step, we weighted the data term using the contour level sets. Weighting the data term is similar to giving scores from 0 to 1 to the least and most informative band of the echo signal power spectrum. On the other hand, considering zero weight does not make sense as there is still power on the less informative parts. Furthermore, assigning zero weights means completely ignoring the data term part of the cost function. Hence, we shifted the calculated $w_d$ and normalized it by dividing it by the maximum value of the new calculated $w_d$. Afterward, we appointed the *L2* regularization norm for attenuation and the *L1* norm for BSC to be in accordance with the physic-based properties of each. Finally, our novel cost function was minimized using ADMM. An important point that should be taken into account is that the expense of the computational complexity for reducing the bias and variance is worth, especially when a phantom or tissue that we attempt to characterize, be similar to the region with specular reflectors meaning that the majority part of the media is with the same QUS parameters and a small fraction of that has different parameters. In such a case, the ALGEBRA completely fails and our proposed technique is highly preferable. But similar to ALGEBRA, a limitation of our method is that for calculating the local attenuation, first, the effective attenuation needs to be estimated. Overall, the best results in terms of bias and variance for both regions of the phantom were acquired by weighting the data term and applying *L1* and *L2* norms in the regularization part of the cost function.

Significant differences between our work and other groups are: I) Proposing weighting the data term. II) Taking into account physics properties of QUS parameter in the cost function.

We predict using $w_d$ and combining *L1* and *L2* norms in the cost function can further improve the bias of estimation for the cases that sample and reference phantoms are different.

## V. CONCLUSION

Herein, we proposed a novel approach for estimating the average attenuation and the BSC. Our algorithm incorporates ADMM to penalize a cost function containing *L2* norm weighted date term and *L1* and *L2* norms regularization terms to be associated with the physical properties of the BSC and the average attenuation. Visual and quantitative evaluations justify that our proposed algorithm substantially outperforms the other techniques.

## ACKNOWLEDGMENT

We acknowledge the support of the Natural Sciences and Engineering Research Council of Canada (NSERC) RGPIN-2020-04612.

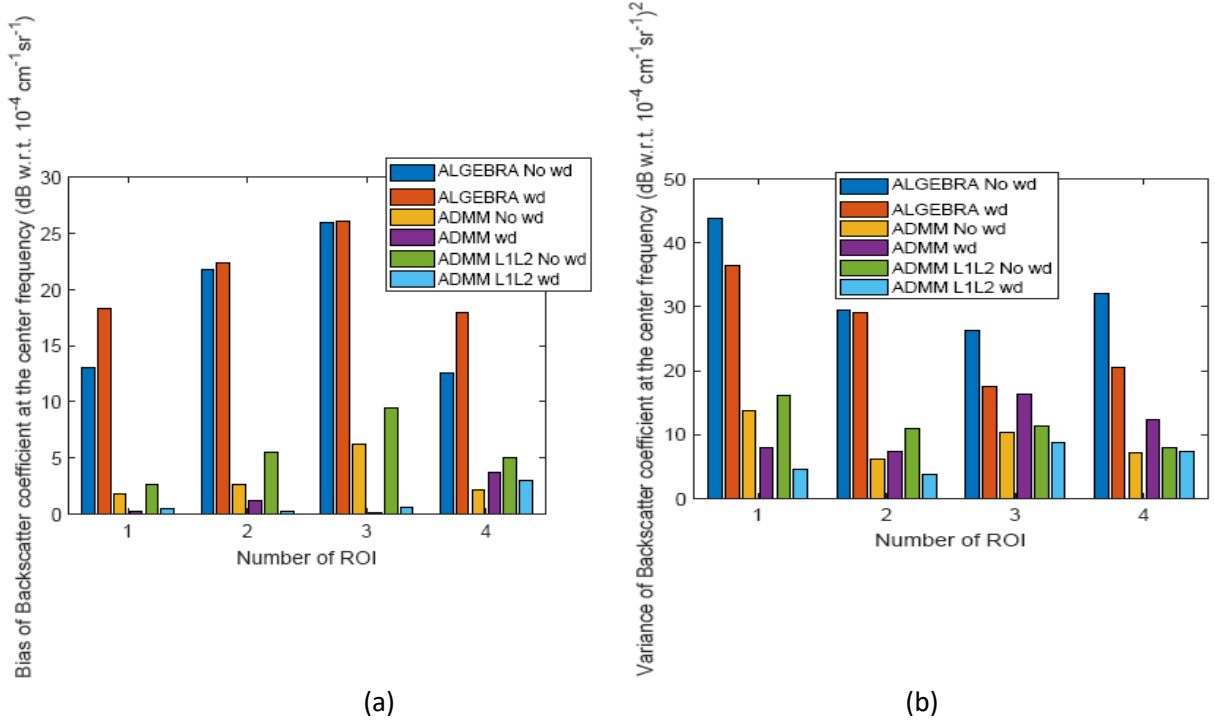

(a) (b)

Fig. 9. Comparison of bias (a) and variance (b) of the BSC using ALGEBRA (ALGEBRA No $w_d$), ALGEBRA associated with weighting the data term (ALGEBRA $w_d$), ADMM without weighting the data term and using *L2* norm in the regularization part for all the parameters (ADMM No $w_d$), ADMM associated with weighting the data term and using *L2* norm in the regularization part for all the parameters (ADMM $w_d$), ADMM without weighting the data term and using *L1* norm for the BSC and *L2* norm for the average attenuation (ADMM L1L2 No $w_d$), and ADMM associated with weighting the data term and using *L1* norm for the BSC and *L2* norm for the average attenuation (ADMM L1L2 $w_d$) in four ROI of the region with specular reflectors. Results are shown on a dB scale with respect to $10^{-4}$ cm$^{-1}$ sr$^{-1}$.

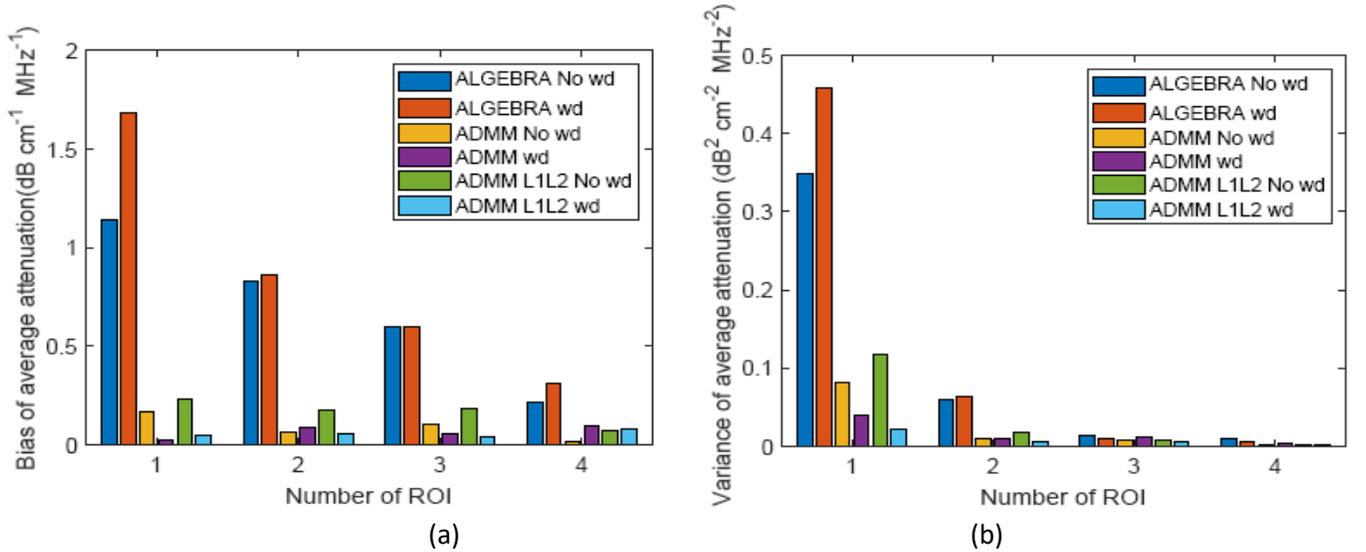

(a) (b)

Fig. 10. Comparison of bias (a) and variance (b) of the average attenuation using ALGEBRA (ALGEBRA No $w_d$), ALGEBRA associated with weighting the data term (ALGEBRA $w_d$), ADMM without weighting the data term and using *L2* norm in the regularization part for all the parameters (ADMM No $w_d$), ADMM associated with weighting the data term and using *L2* norm in the regularization part for all the parameters (ADMM $w_d$), ADMM without weighting the data term and using *L1* norm for the BSC and *L2* norm for the average attenuation (ADMM L1L2 No $w_d$), and ADMM associated with weighting the data term and using *L1* norm for the BSC and *L2* norm for the average attenuation (ADMM L1L2 $w_d$) in four ROI of the region with specular reflectors.

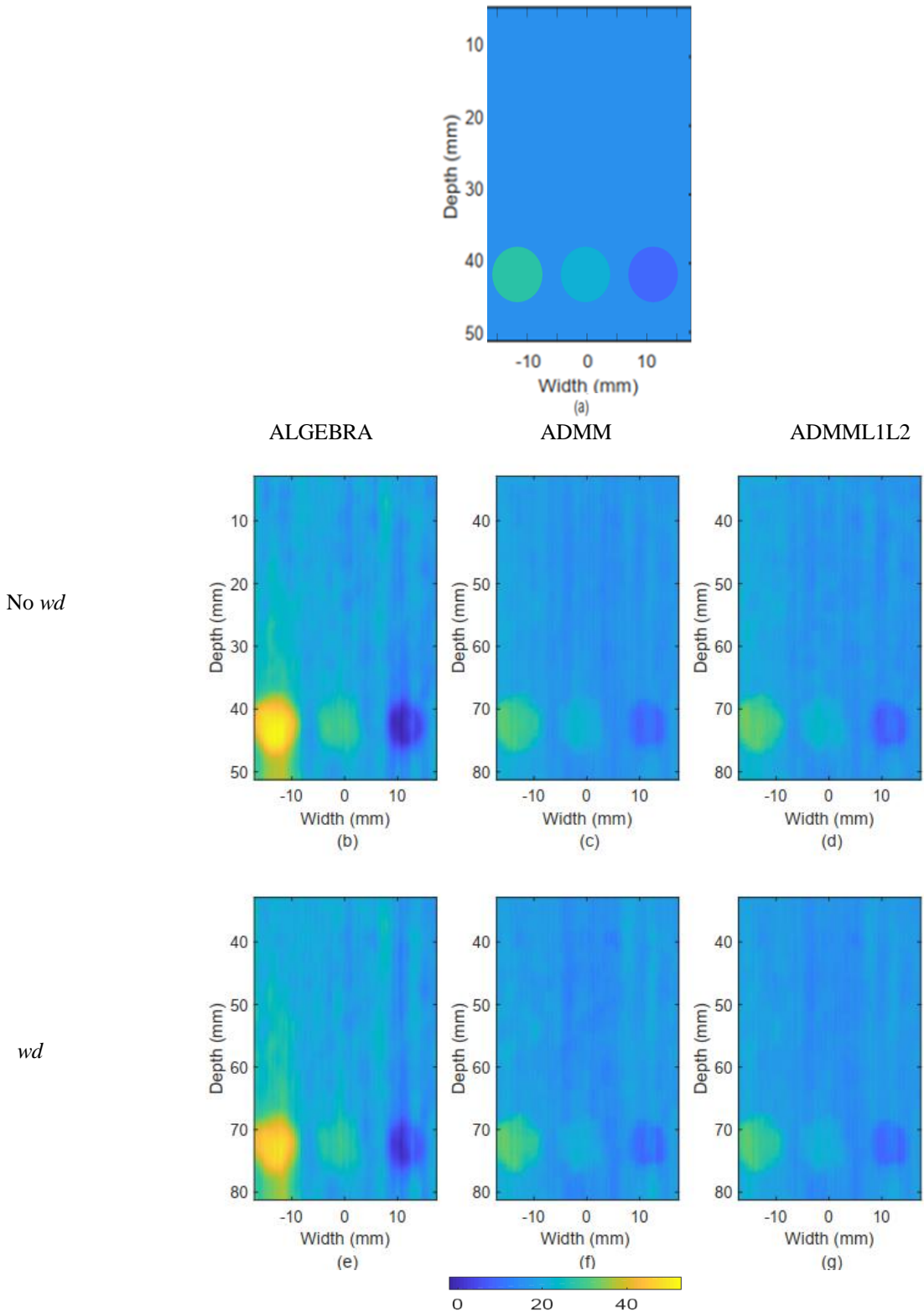

Fig. 11. Parametric image of the BSC at the center frequency of the region with inclusions, GT (a), ALGEBRA (b), ADMM without weighting the data term and using *L2* norm in the regularization part for all the parameters (c), ADMM without weighting the data term and using *L1* norm for the BSC and *L2* norm for the average attenuation (d), ALGEBRA associated with weighting the data term (e), ADMM associated with weighting the data term and using *L2* norm in the regularization part for all the parameters (f), ADMM associated with weighting the data term and using *L1* norm for the BSC and *L2* norm for the average attenuation (g). Results are shown on a dB scale with respect to $10^{-4}$ cm$^{-1}$ sr$^{-1}$.

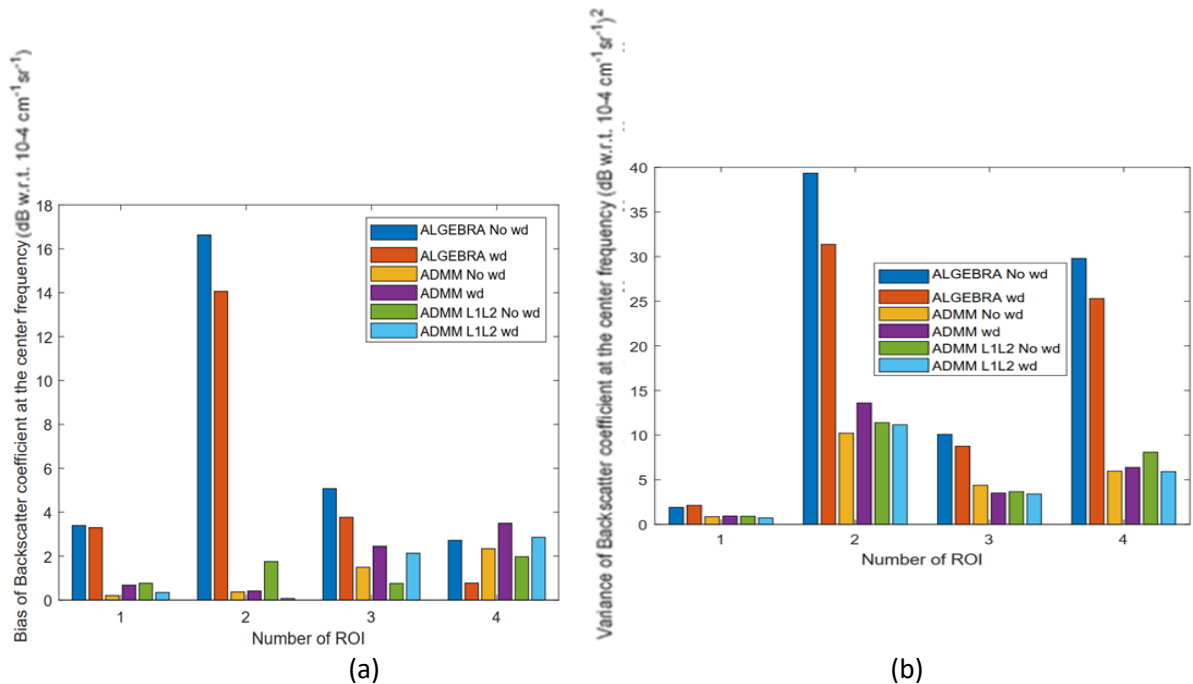

Fig. 12. Comparison of bias (a) and variance (b) of the BSC using ALGEBRA (ALGEBRA No $w_d$), ALGEBRA associated with weighting the data term (ALGEBRA $w_d$), ADMM without weighting the data term and using $L2$ norm in the regularization part for all the parameters (ADMM No $w_d$), ADMM associated with weighting the data term and using $L2$ norm in the regularization part for all the parameters (ADMM $w_d$), ADMM without weighting the data term and using $L1$ norm for the BSC and $L2$ norm for the average attenuation (ADMM L1L2 No $w_d$), and ADMM associated with weighting the data term and using $L1$ norm for the BSC and $L2$ norm for the average attenuation (ADMM L1L2 $w_d$) in four ROI in the region with inclusions. Results are shown on a dB scale with respect to $10^{-4}$ cm$^{-1}$ sr$^{-1}$.

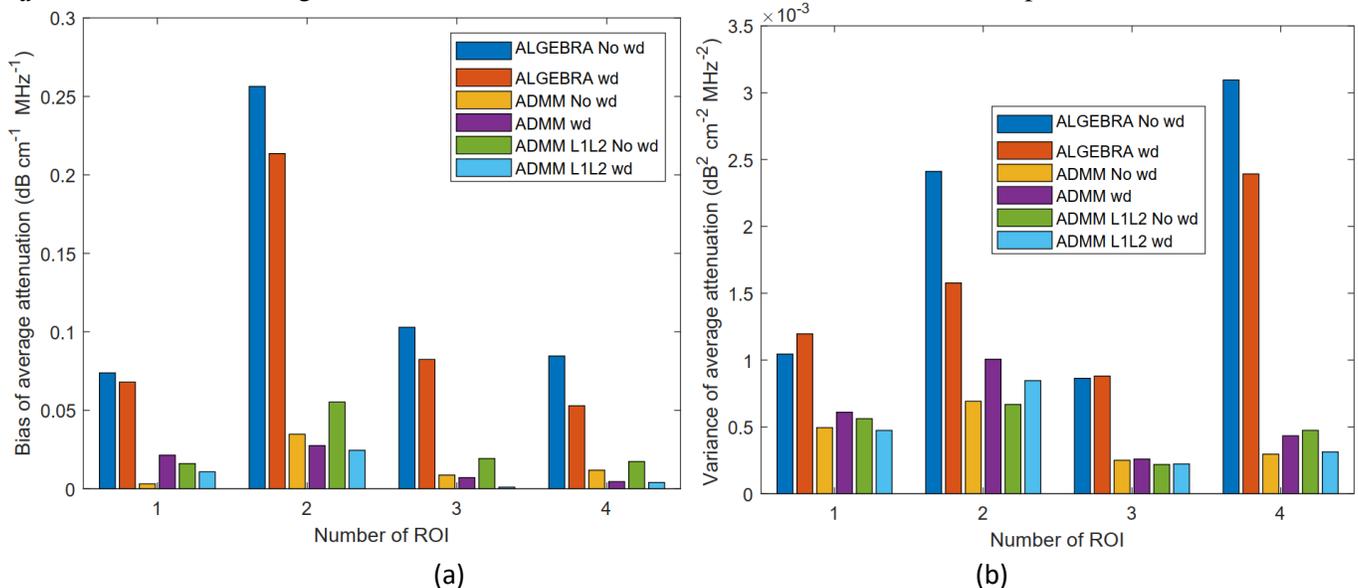

Fig. 13. Comparison of bias (a) and variance (b) of the average attenuation using ALGEBRA (ALGEBRA No $w_d$), ALGEBRA associated with weighting the data term (ALGEBRA $w_d$), ADMM without weighting the data term and using $L2$ norm in the regularization part for all the parameters (ADMM No $w_d$), ADMM associated with weighting the data term and using $L2$ norm in the regularization part for all the parameters (ADMM $w_d$), ADMM without weighting the data term and using $L1$ norm for the BSC and $L2$ norm for the average attenuation (ADMM L1L2 No $w_d$), and ADMM associated with weighting the data term and using $L1$ norm for the BSC and $L2$ norm for the average attenuation (ADMM L1L2 $w_d$) in four ROI in the region with inclusions.